\documentclass{article}
\usepackage[centertags]{amstex}
\usepackage{amssymb}
\usepackage{epsfig}		

\makeatletter
\renewcommand\section{\@startsection{section}{1}{\z@}%
                                   {-3.5ex \@plus -1ex \@minus -.2ex}%
                                   {2.3ex \@plus.2ex}%
                                   {\reset@font\Large\scshape}}
\renewcommand\subsection{\@startsection{subsection}{2}{\z@}%
                                     {-3.25ex\@plus -1ex \@minus -.2ex}%
                                     {1.5ex \@plus .2ex}%
                                     {\reset@font\large\slshape}}
\makeatother
\newcommand\nablaright{\overrightarrow{\nabla}}
\newcommand\nablaleft{\overleftarrow{\nabla}}
\newcommand\Tr{\operatorname{Tr}}
\newcommand\bbone{1\kern-3.9pt 1}
\newcommand\amod[1]{\lvert #1\rvert}

\begin{document}

\thispagestyle{empty}

\begin{flushright}
\large
SWAT/51 \\
hep-lat/9412065
\end{flushright}

\vspace{1cm plus 1fil}

\begin{center}

\textbf{\huge Point-to-point Hadron Correlation Functions using the
Sheikholeslami--Wohlert Action}

\vspace{1cm plus 1fil}

{\Large UKQCD Collaboration}

\vspace{0.5 cm}
{\large
{\bfseries S.J. Hands and P.W. Stephenson \\}
           Department of Physics, \\
           University of Wales, Swansea, \\
           Singleton Park, Swansea, SA2 8PP, U.K.\\

\vspace{0.3cm}

                       and \\

\vspace{0.3cm}

{\bfseries A. McKerrell \\}
           DAMTP \\
           University of Liverpool, \\
           Liverpool, L69 3BX, U.K.

}

\vspace{1cm plus 1fil}

\textbf{\Large Abstract}

\end{center}

\noindent 
We calculate correlations between hadronic current operators as a
function of their spatial separation, in a quenched lattice QCD
simulation at $\beta=6.2$ on a $24^3\times48$ lattice. The lattice
fermion formulation used is that due to Sheikholeslami and Wohlert.
The correlation functions are then compared with the corresponding
quantities calculated in the infinite volume chiral limit of the
non-interacting theory. The ratio of the two quantities contains
information on the vacuum structure of QCD. Results obtained are
consistent with previous studies, and with known hadron phenomenology,
although certain features of the fermion action used make it impossible
to probe the $r\to0$ limit directly, thus limiting the accuracy possible.

\newpage

\section{Introduction}

In lattice QCD simulations, huge amounts of numerical data are
generated in the form of correlation functions between field operators.
The standard means of analysis is first to project onto a particular
momentum sector (for example by summing over spatial sites to construct the 
``timeslice propagator'' with ${\bf p}=0$), and then to study the decay
of the correlator for large time separations. If the smallest eigenvalue
of the transfer matrix is $M$, then the asymptotic form for the decay is
proportional to $\exp(-Mt)$. Whilst this is by far the most effective way of
extracting spectroscopic information, one might wonder if more use could
be made of correlation data from short and intermediate distance
separations, which is in effect either discarded or integrated over.
Is it always necessary to focus on the asymptotic regime?

In a recent review~\cite{Shu93}, Shuryak has argued that there is much
more information available, from a consideration of point-to-point
correlation functions between hadronic current operators defined using
quark fields. In the Euclidean formalism, the basic object is
\begin{equation}
K(x)=\langle0\mid J(0)\bar J(x)\mid0\rangle, \label{eqn:J}
\end{equation}
where $J$ is a current operator with spacetime and spin structure
appropriate to the channel being discussed: e.g.\ for the vector current
correlator we define
\begin{equation}
J_{\mu}(x)=\bar q(x)\gamma_\mu q(x),
\end{equation}
and then contract over the index $\mu$ in evaluating equation~(\ref{eqn:J}),
in order to obtain a scalar function. Note that we assume
translational invariance, so that one current operator sits at the
origin; a further assumption of Euclidean rotational invariance
enables us to replace the displacement vector $x$ by $r=\lvert
x\rvert$.  The correlator $K(x)$ is a very natural object to measure
in lattice simulations, either of the quenched or the full theory, and
first results from such a study have already appeared~\cite{Chu93}.

In practice, to avoid unwieldy kinetic factors, and to
facilitate comparison between different channels, it is useful to
consider the ratio between the correlation function in the interacting
theory and that calculated for non-interacting massless quarks:
\begin{equation}
R(x)\equiv{K_{\rm interacting}(x)\over K_{\rm free}(x)}.
\end{equation} 
Since QCD is asymptotically free, one can immediately see
qualitatively that in the limit $r\to0$ this ratio should approach
unity; the interaction must be weak at short
distances. Conversely, whereas for the free theory we expect $K_{\rm
free}(r)$ to decay as a power of $r$ as $r\to\infty$, in the
interacting theory we expect the lightest state (with the exception of
the pseudoscalar channel, which is dominated by a Goldstone mode) to
have a mass of the order of $\Lambda_{\rm QCD}$, and hence $K_{\rm
interacting}(r)\sim\exp(-{\rm const}\times\Lambda_{\rm QCD}r)$ for
large $r$. Hence $R(r)$ tends to zero as $r$ tends to infinity.

\begin{figure}[tb]
\begin{center}
\psfig{file=fitparams.eps,width=3in}
\end{center}
\caption{Parameters $m$, $\lambda$ and $s_0$ and the phenomenological
fit for the spectral density $f(s)$.}
\label{fig:spec}
\end{figure}

The real power of the approach arises because the Fourier transform of
$K(x)$ can be related via a dispersion relation to the spectral density
function $f(s)$~\cite{Shu93,Chu93}:
\begin{equation}
\int d^4x e^{iqx}K(x)=P(q)\int ds {{f(s)}\over{s+q^2}}+Q(q^2), \label{eqn:fdef}
\end{equation}
where $P(q)$ is a kinematic factor whose details depend simply on
the channel under consideration, and $Q(q^2)$ is a finite-order
polynomial, whose only effect in $x$-space is to generate a contact
interaction at $r=0$. Thus $K(x)$ and $f(s)$ contain essentially
equivalent information. In general $f(s)$ will have contributions both
from the lowest lying resonance in a given channel (e.g.\ the rho meson
in the vector channel) as well as higher mass resonances and
multi-particle continua (the latter is only expected in the full
theory, not the quenched).  For many practical purposes it is
convenient to parametrise $f(s)$ as follows~\cite{Chu93}:
\begin{equation}
f(s)=\lambda^2\delta(s-M^2)+f_c(s)\theta(s-s_0). \label{eqn:fform}
\end{equation}
Here $M$ is the mass of the lowest lying resonance, assumed to have zero
width (this should be exact in the quenched approximation),
$\lambda$ is the coupling of the
hadronic current to that particular state, and all higher mass states
have been parametrised by a continuum $f_c(s)$ starting at some
threshold $s_0$. A further approximation, justified by asymptotic
freedom as $s\to\infty$, is that $f_c(s)$ is replaced by the free-field
spectral density $f_0(s)$, which can be found
analytically~\cite{Chu93}.  This is shown schematically in
figure~\ref{fig:spec}.

We will use the form~(\ref{eqn:fform}) in all the analyses presented
here. However, it is worth stressing that in principle $f(s)$ as
determined from equation~(\ref{eqn:fdef}) can be much more complicated
than the ansatz~(\ref{eqn:fform}), and that in particular there is no
reason {\sl a priori\/} to expect the replacement of $f_c(s)$ by
$f_0(s)$ to be accurate, since perturbation theory is still not
trustworthy at the values of $s_0$ found, typically $\mathcal{O}(1{\rm GeV})$.
Shuryak~\cite{Shu93} has used $f(s)$ derived from phenomenological
scattering data to reconstruct $K(x)$: e.g.\ for the $I=1$ vector
channel $f(s)$ can be extracted from the cross-section of
$e^+e^-\longrightarrow N\pi$, where $N$ is even. Thus direct
comparison of $R(x)$ as calculated on the lattice and as extracted
from phenomenology is possible. We refer the reader throughout to the
original references~\cite{Shu93,Chu93}.

Apart from the opportunity to compare numerical predictions with QCD
phenomenology in a hitherto under-explored regime, point-to-point
correlation functions may be of intrinsic interest to lattice theorists.
They offer a further chance to compare different lattice fermion
formulations; these might be expected to differ by effects which are
$O(a)$, $O(a^2)$ etc, where $a$ is the lattice spacing, and hence which
manifest themselves at short distances, rather than the asymptotic
separations necessary for spectroscopy. It may be interesting to study 
the crossover behaviour from the perturbative regime (where corrections to,
say, the Wilson fermion formulation are well understood and under control) to
the non-perturbative, where our knowledge about any lattice fermion
scheme is more limited. Most importantly, it may well prove to be the
case that the differences between quenched and full QCD simulations will
show up not in spectroscopy, where relatively heavy dynamical quarks
have had little impact to date~\cite{Kro93}, but in the short distance behaviour
governed by the higher resonances.

As found in~\cite{Chu93}, the analysis of point-to-point correlation
functions involves the treatment of systematic effects which have not
been studied in detail up to now. Hence it will be necessary to study
the problem using a variety of fermion formulations, lattice spacings,
and physical volumes in order to gain a controlled understanding.  The
original study used quenched QCD data generated with a physical
lattice spacing $a=0.17{\rm fm}$ and a spatial lattice size of 2.7fm.
The fermion formulation used was the Wilson scheme. In this paper we
will use data taken by the UKQCD collaboration~\cite{All93} with
lattice spacing $a=0.07{\rm fm}$ and spatial lattice size 1.8fm.  Thus
we have information at a much finer cutoff, but in a slightly smaller
physical volume. The biggest difference is our use of the fermion
formulation first proposed by Sheikholeslami and Wohlert~\cite{She85}
--- which we will refer to as the SW formulation --- which eliminates
the $O(a)$ corrections which occur in perturbative calculations using
the Wilson formalism \cite{Hea91}, by the introduction of a term of
the form $g^2\sigma_{\mu\nu}F_{\mu\nu}$ into the standard formulation,
together with a redefinition of the quark fields in the functional
integral. As we shall see, the SW formulation has rather unexpected
properties, possibly attributable to the presence of contact terms
induced by the field redefinition, which make it difficult for us to
examine the limit $r\to0$. Further details of the generation of the
gauge configurations and calculation of the SW propagators is given in
ref.~\cite{All93}. Of course, for calculation of (1), it is necessary
to use local sources for the propagator inversions, and so we have
made no use of smearing.

\section{Analysis}

\begin{table}[tb]
\renewcommand{\arraystretch}{1.3}
\begin{tabular}{lr@{$\>=\>$}lc}
\hline
\multicolumn{1}{c}{Channel} & \multicolumn{2}{c}{Current} &
  Correlator $K(x)$ \\
\hline\hline
Pseudoscalar & $J^5$ & $\bar u\gamma_5 d$ & 
  $\langle J^5(x)\bar J^5(0)\rangle$ \\
Vector & $J_\mu$ & $\bar u\gamma_\mu d$ & 
  $\langle J_\mu(x)\bar J_\mu(0)\rangle$ \\
Axial & $J^5_\mu$ & $\bar u\gamma_\mu\gamma_5 d$ & 
  $\langle J^5_\mu(x)\bar J^5_\mu(0) \rangle$ \\
Scalar & $J^s$ & $\bar ud$ & $\langle J(x) \bar J(0) \rangle$ \\
Nucleon & $J^N$ & $\epsilon_{abc}[u^aC\gamma_5 d^b]u^c$ & 
 $\frac{1}{4} \Tr [\langle J^N(x)\bar J^N(0)\rangle x_\mu\gamma_\mu]$ \\
\hline\hline
\end{tabular}
\caption{Definitions of the currents and the point-to-point
correlators.}
\label{tab:ops}
\end{table}

Much of our analysis is inspired by Chu et al.~\cite{Chu93}, which
contains the first detailed report on point-to-point correlations in
lattice gauge theory.  We refer the reader to that paper for details
concerning the exact forms for fits to phenomenological predictions,
and so on; here we simply summarise where appropriate.  The
definitions of our operators and correlators are shown in
table~\ref{tab:ops}; the only difference from ref.~\cite{Chu93} is in
our definition of the nucleon current which agrees with the convention
usually used by UKQCD. Note that the definition of the nucleon
correlator $K_N(x)$ incorporates a contraction with the matrix
$x{\!\!\! /}\,$: as well as defining a scalar function, this maximises
the use of data from the various spinor components of the nucleon
current.

Our data shows larger lattice artefacts at small distances than that
of ref.~\cite{Chu93}, which has necessitated some additional cuts on the
data.  Also, we have performed a different extrapolation to the chiral
limit.  Naturally, we hope that the same physics will appear through
the differences.

\subsection{Data}

Our data is derived from quenched SU(3) configurations on an $L^3\times
T = 24^3\times48$ lattice at inverse coupling $\beta=6.2$.  Three quark
masses were available to us, those corresponding to hopping parameters
$\kappa=0.14144$, $0.14226$ and $0.14262$ using the SW action.  Only
27 configurations were available at all three values and we have
restricted our analysis to these for consistency.  The mass spectrum
of this data was analysed in ref.~\cite{All93}; the hopping parameter
corresponding to a massless pseudoscalar was found by extrapolation to
be $\kappa_\mathrm{critical} = 0.14313(+7-4)$.  Our own extrapolation to
the chiral limit using this value is discussed in detail below.

The first step in the procedure is to generate the raw correlators.
The definitions of the currents $J(x)$ used mean that the correlators
for mesons are symmetric about the mid-point of the lattice in each of
the three spatial directions, while those for baryons are
anti-symmetric.  This allows us to use data from all points on the
spatial lattice in the former case, while we reject points with any
site with a spatial co-ordinate $L/2$ (i.e.\ 12) in the latter case.
We have examined propagators from the origin to times 0 and 1.  For
time 1, we have verified directly that the lack of rotational
invariance in the time direction due to the asymmetric lattice volume 
is insignificant compared with the
statistical errors in the data.

All points on all data sets which are exactly equivalent by rotation
or reflection are combined into the same data point, with errors
generated by a bootstrap of all the data.


\subsection{Correction for finite volume effects}

Our goal is to look at strong-interaction physics over a range of
different length scales.  This means that we need to have as much
control as possible over the data both at short and long distances.

The finite volume effects at long distance are dominated by the images
of the delta function source at the origin against which the
propagators are inverted; these images occur at all points
$\mathcal{L}_0$ on the simple cubic lattice which are translations of
the origin by $L$ in $x$, $y$ or $z$.  The effects become significant
when one co-ordinate $x_i$ becomes near enough to $L-x_i$ for the
propagator from the nearest source (which is only the origin if all
spatial $x_i$ are less than $L/2$) to be comparable in size to that
from the the image in $x_i=L/2$.  As spatial points may be close to
none, one, two or three of the $x_i = L/2$ planes, this shows up as
clear anisotropies in the data.

A procedure exists for removing these effects~\cite{Chu93,Bur94}.  The
idea is to model the tail of the data by its asymptotic value; e.g.\ for
mesonic channels we assume the form for a free boson of mass $M$:
\begin{equation}
K_\mathrm{interacting}(r) = Cr^{-3/2}e^{-Mr} \label{eqn:correct}
\end{equation}
The original data is then assumed to be the sum of the ``true'' signal
plus contributions from all the images of the tail (6).
This works on the assumption that the physical volume is large enough
such that for all elements of $\mathcal{L}_0$ except
the nearest only the asymptotic part of the correlation is
significant; note also that contributions in which, say, the quark
interacts with the image of the anti-quark are strongly suppressed by
the confinement mechanism. This is not the case for the free correlators
$K_{\rm free}(r)$, discussed below.

We fit the mass $M$ and the coefficient $C$ to the raw data and use
the fitted function to subtract the effects of the images.  This is
iterated until a self-consistent result for the image-corrected signal 
is obtained.  The procedure
appears to be completely successful:  there is no remaining sign of
anisotropy from the long-distance effects.  We also note that the
masses obtained from the fit in equation~(\ref{eqn:correct}) compare
well with those obtained by phenomenological fits to the data
separately at each quark mass (not reported here).




\subsection{Correlations from non-interacting quarks}

Next we detail the calculation of the free fermion propagator on a
lattice, which is needed to construct free hadron correlation functions
to normalise the results from the interacting case. One might hope that
in taking the ratio in this way some of the lattice anisotropy effects
will be cancelled; enough remain, however, to need attention, as
discussed below.

Firstly we should explain why this needs to be done using a different
lattice size and hopping parameter to those used to generate the
interacting theory data. The reason is that systematic effects from
image and non-zero mass corrections are qualitatively very different
for free and interacting theories, and hence we must attempt to
extrapolate each case to the limits in which $V\to\infty$ and bare
quark mass $m\to0$ independently. Fortunately, in the free case there
is a semi-analytic prescription for free fermion propagators due to
Baillie and Carpenter~\cite{Car85}, which makes it possible to attain
these limits for the range of $r$ under consideration almost
perfectly.  We start from the Wilson fermion action (with Wilson
parameter $r=1$) defined on a $L^3\times T$ lattice:
\begin{multline}
S_W=\sum_{x\mu}{1\over2}[\bar\psi(x)(\gamma_\mu-\bbone)\psi(x+\hat\mu)-
                           \bar\psi(x+\hat\mu)(\gamma_\mu+\bbone)\psi(x)] \\
+(m+4)\sum_x\bar\psi(x)\psi(x).
\end{multline}
Here $m$ is the bare quark mass in the classical continuum limit. The
propagator is first generated in 3-momentum space and real time using
the decomposition
\begin{equation}
S({\bf k},t)=\sum_iS_i({\bf k},t)\gamma_i+S_4({\bf k},t)\gamma_4
+S_u({\bf k},t)\bbone,
\end{equation}
where $i$ runs over the three spacelike directions, and
\begin{align}
S_i({\bf k},t)&=-i{{\sin k_i}\over{2\mathcal{E}({\bf k})}}
\left[{{e^{-E\amod{t}}+be^{-E(T-\amod{t})}}
\over{1-be^{-ET}}}\right];  \\[4pt]
S_4({\bf k},t)&={\rm sgn}(t){{\sinh E({\bf k})}\over{2\mathcal{E}({\bf k})}}
\left[{{e^{-E\amod{t}}-be^{-E(T-\amod{t})}}
\over{1-be^{-ET}}}\right];  \\[4pt]
S_u({\bf k},t)&={{1-\cosh E({\bf k})+m+\sum_i(1-\cos k_i)}
\over{2\mathcal{E}({\bf k})}}
\left[{{e^{-E\amod{t}}+be^{-E(T-\amod{t})}}
\over{1-be^{-ET}}}\right]  \notag\\
&\quad+\delta_{t0}{1\over{2(1+m+\sum_i(1-\cos k_i))}}; 
\end{align}
with
\begin{align}
E({\bf k})&=\cosh^{-1}\left(1+{{\sum_i\sin^2k_i+(m+\sum_i(1-\cos k_i))^2}
\over{2(1+m+\sum_i(1-\cos k_i))}}\right); \\
\mathcal{E}(\mathbf{k})&=(1+m+{\textstyle\sum_i}(1-\cos k_i))\sinh E({\bf
k});
\end{align}
and
\begin{equation}
\mathrm{sgn}(t)\equiv{t\over{\amod{t}}}\;;\qquad\mathrm{sgn}(0)
\equiv0.
\end{equation}
The constant $b=\pm1$ depending on whether periodic or antiperiodic
boundary conditions are applied in the time direction. The lattice
3-momenta $k_i$ are given by
\begin{equation}
k_i={{2\pi}\over L}(m+c),\qquad m=0,\ldots,L-1,
\end{equation}
where $c=0$ for spatial periodic boundary conditions and $c={1\over2}$
for antiperiodic boundary conditions.

Next, the real space propagator $S(x)$ is constructed using a three
dimensional Fourier transform. Finally, in order to compare with
interacting quark propagators generated using the SW action, the field
redefinitions must be implemented by rotating the propagator at either
end~\cite{Hea91,All93}:
\begin{equation}
S(x-y)\equiv S(x,y)\mapsto S_R(x,y)=
(1+{\textstyle{1\over2}}{\nablaright}_{\mu x}\gamma_\mu)S(x,y)
(1-{\textstyle{1\over2}}\nablaleft_{\nu y}\gamma_\nu)
\label{eqn:freeprop}
\end{equation}
where
\begin{equation}
\nablaright_{\mu
x}f(x)\equiv{1\over2}\left(f(x+\hat\mu)-f(x-\hat\mu)\right).
\end{equation}
Note that because we are working in the free field limit, the rotation
(\ref{eqn:freeprop}) is the only distinction between Wilson and SW
formulations. The
hadron correlation functions can now be constructed in the channels of
interest, using the definitions in table~\ref{tab:ops}. 
Taking advantage of translational invariance, and defining
\begin{align}
G(x)&=\sum_\mu S^2_{\mu R}(x);\\
M(x)&=S^2_{uR}(x),
\end{align}
we find for the hadron correlators in the various channels:
\begin{equation}
\begin{split}
K_V(x)&=8(G(x)+2M(x)) \\
K_A(x)&=8(G(x)-2M(x)) \\
K_P(x)&=4(G(x)+M(x))  \\
K_S(x)&=4(-G(x)+M(x)) \\
K_N(x)&=4(5G(x)+7M(x))\times\sum_\mu x_\mu S_{\mu R}(x)
\end{split}
\end{equation}
For all channels we examined our results for the free propagators were
tested against propagators extracted using a conjugate gradient routine
on a lattice with unit link variables.

We chose to employ antiperiodic boundary conditions in all directions,
for two reasons. First, since there is no zero momentum mode, the bare
mass $m$ can be set equal to zero, and we can work directly in the
chiral limit. Secondly, finite volume effects are thus minimised.  The
chief source of such effects turns out to be the contribution
proportional to the unit matrix in spin space, i.e.\ $S_{uR}(x)$. This is
an even function of $x$, and so contributions from paths which wrap one
or more times round the lattice will add up if periodic boundary
conditions are used. Since in the $m\to0$ limit we expect a power-law
fall off for $S_R(x)$, then contributions from essentially an infinite
number of lattice circuits count, so that $S_{uR}(x)$ is very nearly
independent of $x$ as $m\to0$ on a periodic lattice. If antiperiodic
boundary conditions are used, then the wrapping corrections form an
alternating series whose sum is numerically much less significant.

\begin{figure}[tb]
\psfig{file=free.ps,width=\textwidth}
\caption{Comparison of the continuum free correlator in the vector
channel evaluated using
massless quarks, with the lattice free correlator on a $96^4$ lattice
described in the text.}
\label{fig:free}
\end{figure}

In figure \ref{fig:free} 
we show the results for the free correlator in the vector
channel calculated on a $96^4$ lattice for lattice separations 
$r$ in the range 16 -- 21, which is the largest separation we
can sensibly discuss using our interacting propagators obtained on a
$24^3\times48$ lattice. Also shown is the theoretical expectation for
free massless fermions: $K_V(r)={2\over\pi^4}r^{-6}$. The agreement at
the furthest point is good to 2\%; the error must arise from a combination of
finite volume and discretisation errors. We also checked against the same
quantity computed on a $84^4$ lattice; the agreement was to within 
0.3\%, comparable with the naive expectation for finite volume
corrections at this value of $r$ of $O(12/96)^3$.
Of course, any systematic error for the free massless correlation
function is dwarfed by the statistical error for the interacting
correlators (described below). For all practical purposes the free
propagators we use are those of the infinite volume chiral limit.

\subsection{Finite lattice spacing effects}

The remaining major contribution to lattice artefacts, from finite lattice
spacing effects, is harder to deal with.  The authors of
ref.~\cite{Chu93} found using Wilson fermions that points $x$ nearest the
Cartesian axes produced anomalously large correlations in the
interacting data, in accordance with the na\"\i ve expectation that
those points, having the fewest contributions in a hopping parameter
expansion, would show the greatest artefacts.  Their ansatz was to cut
the data to points near the body-diagonals of the spatial lattice; 
attention was restricted to a
cone of points within an angle $\theta$ such that $\cos\theta > 0.9$.

\begin{figure}[tb]
\psfig{file=zeroes.ps,width=\textwidth}
\caption{Anisotropy effects in the vector channel for the heaviest
quark, $\kappa=0.14144$; the horizontal
scale shows the distance in lattice units while the vertical scale
is arbitrary.}
\label{fig:anis}
\end{figure}

In our case anisotropy persists out to some eight lattice spacings,
and towards the $r=8a$ end the appearance is similar to that in the
Wilson case.  It is to be expected that anisotropy effects for SW
fermions will manifest themselves most at small distances.  This is
what we indeed observe: consider the vector channel for convenience,
where the expectation~\cite{Shu93} is that $R(r)$ remains roughly
constant at small distances.  As $r$ goes to zero, the correlators
appear to reach some threshold and then plummet.  For definiteness, we
categorise the points by the number of zeroes in the cartesian
co-ordinates $(x,y,z,t)$.  This is shown in figure~\ref{fig:anis}
for the largest quark mass, $\kappa=0.14144$, where the
effect is clearest.  It
appears that the points nearest the diagonals, i.e.\ the $t=1$ points
with $x$, $y$ and $z$ non-zero, in fact have the most pronounced
behaviour in this respect.  This effect was not remarked upon by Chu
et al.~\cite{Chu93} and is presumably specific to the SW action. One
possible explanation is that the apparently singular behaviour is due
to contact terms induced by the rotation (16) necessary to perform the
field redefinition in the SW formalism~\cite{Hea91}. We should like
to have a further analytic understanding of this point.

In practice this means that we cannot trust our results for $r\lesssim 6$,
even with the $\cos\theta$ cut.  These points are included in
subsequent diagrams, but play no part in our fits.  We hope to
elucidate the differences between Wilson and SW fermions in more
detail in future work.

\subsection{Chiral extrapolations}

There are three obvious possibilities for extrapolating to the chiral
limit at $\kappa_\mathrm{critical}= 0.14313$, namely extrapolating
linearly by fitting $R(\kappa, x) = A(x) + B(x)
(1/\kappa-1/\kappa_\mathrm{critical})$, adding a quadratic term to the
linear fit, or performing a linear fit to $\log R(x)$.  We find that a
simple linear extrapolation gives a result significantly different
from the other two, which are quite similar.  Therefore we have chosen
the fit logarithmic in $R(x)$ as it has fewer parameters.

Our points at large distances have large error bars and the errors will
be magnified in the extrapolation.  We have therefore investigated
ways of controlling these errors while retaining the shape of the data
which is where much of the interest lies.

As we perform the procedure separately at each point, our fit is really
\begin{equation}
\log R(\kappa, r) = R_\mathrm{c}(r) + B(r)\left({1\over\kappa} -
{1\over\kappa_\mathrm{critical}}\right)
\label{eqn:logfit}
\end{equation}
where the fit parameters are assumed to be
functions of $r$; $R_\mathrm{c}(r)$ is
the extrapolated value and $B(r)$ the slope of the extrapolation.
Inspection of the data confirms that not only $R_\mathrm{c}$, 
but also $B$, are smooth functions of $r=\lvert x\rvert$.
Smoothness of $B(r)$ is required in the following. The
magnification of the errors by the fit arises from fluctuations of $B$ 
from its correct value.  Therefore, we choose to block values of $B$
taken from this fit together into bins of width $2a$ and hence make
the values of $B$ used smoother.  We then interpolate linearly between
the central points of the bins to obtain a smoothed $B(r)$: note that
the lack of smoothness in the slope of $B$ itself does not have any
effect other than a small increase in the errors.  This smoothed
$B(r)$ is fed back into equation~(\ref{eqn:logfit}) and a second, single
parameter, fit performed for $R_\mathrm{c}(r)$.

We have the problem of deciding what are the errors on the values
$R(r)$ (dropping the subscript) obtained.  If the procedure is
effective --- as it appears to be --- we would hope that the errors
have been reduced.  Thus we consider it safe to use the errors on the
original (two parameter) fit rescaled to the new data point.  In fact,
these resulting errors are not used for any quantitative purpose
beyond the fits described in the next section (and we do not quote the
errors on the fit parameters in table \ref{tab:fit} from this source).  
Furthermore, it is
clear that there are strong correlations in the data for which
there is no simple way to account; the block-smoothed values will
merely add to the existing correlations. Hence we do not consider this
point as being of fundamental importance.

We find that, indeed, the block-smoothed extrapolations are less
wayward than the na\"\i ve ones while retaining the same shape of the
curve.  We repeat that our chief interest is in the shape of the
correlations, which this method, despite the added complexity,
retains; naturally, we have also checked that no spurious features are
introduced.  We have therefore used these values in the subsequent
analysis.  Clearly, however, data from more quark masses would be very useful
for this type of work where the chiral extrapolation is problematic.

\subsection{Results and fits to phenomenological forms}

As discussed in the introduction, 
the central quantity in the phenomenology of the point-to-point
correlations is the spectral density, $f(s)$. 
It can be used to calculate the ratios $R(r)$ by
transforming to real space with the propagator for a particle of
energy $\sqrt{s}$ to distance $r$, $D(\sqrt{s}, r)$,
\begin{equation}
R(r) = {1\over\pi} \int ds\,f(s) D(\sqrt{s},r),
\end{equation}
which for the simple $f(s)$ we use produces an expression involving
modified Bessel functions~\cite{Chu93}.

The spectral density therefore includes all information about
resonances and non-resonant contributions to the channel and can in
principle be extremely complicated.  We follow ref.~\cite{Chu93} in
including one delta-function, as shown above in
equation~(\ref{eqn:fform}) and figure~\ref{fig:spec}.  In addition,
there is one extra fit parameter $C_n$ corresponding to an overall
normalisation (which, as we have shown, cannot be extracted from the
short distance limit as one might wish). In fact, the factor $C_n$
also absorbs the perturbative renormalisation factor, in principle
calculable, which should be applied to relate the lattice hadronic
current operator to its continuum counterpart. Here we assume that the
factor applies equally to both resonance and continuum contributions.

We have performed the fits for the pseudoscalar, vector and nucleon
channels; the forms are the same ones given in ref.~\cite{Chu93}.  We
limit the fit to $r>6a$ where the results appear to be trustworthy.
This is probably too conservative, but the points for small $r$ have
small errors and hence dominate the fit, so it is important to ensure
spurious effects are suppressed in this region.  The points at large
distance have sizeable errors, but their contribution to the $\chi^2$
is so small (possibly too small, as we shall argue below) that there
is no good reason to make data cuts in this region.

The errors on the fit parameters derive from those on the data points
which as we have mentioned have large correlations and are not
expected to be accurate.  For a better assessment, we use a global
jackknife of all the data: each data set is eliminated in turn and the
entire analysis from image correction through to chiral extrapolation
is repeated on each subset.

\newlength\figwidth
\setlength\figwidth{\textwidth}
\addtolength\figwidth{-1cm}
\begin{figure}[!p]
\psfig{file=pseudofits.ps,width=\figwidth}
\caption{The pseudoscalar data (only for $t=0$ for clarity) for $R(x)$
and a four parameter fit to it. As for all the fits shown, the
normalisation of the data is arbitrary.}
\label{fig:pseudo}
\end{figure}

\begin{figure}[!p]
\psfig{file=vectorfits.ps,width=\figwidth}
\caption{The vector data for $R(x)$ at $t=0$ and two fits:  the solid
line is a full four parameter fit while the dashed line has the
resonance mass fixed to the value extracted by standard methods.}
\label{fig:vector}
\end{figure}

\begin{figure}[!p]
\psfig{file=nucleonfits.ps,width=\figwidth}
\caption{The data in the nucleon channel, shown as in
figure~\ref{fig:vector}.}
\label{fig:nucleon}
\end{figure}

\begin{figure}[!p]
\psfig{file=axialfits.ps,width=\figwidth}
\caption{The data in the axial channel with a three parameter fit.}
\label{fig:axial}
\end{figure}

\begin{figure}[t]
\psfig{file=scalarfits.ps,width=\figwidth}
\caption{The data in the scalar channel with a single parameter fit to
the normalisation and the continuum threshold fixed at a small value.}
\label{fig:scalar}
\end{figure}

Most of the resulting errors are quite acceptable (of order 10\%) but
those for the continuum cut-off $s_0$ are large.  This is scarcely
surprising considering that the continuum effects are largest at small
$r$ where we are unable to fit.  In particular, this
renders the fit much less stable for the nucleon where the errors in
the nucleon mass rise to nearly fifty percent.  However, the jackknife
procedure seems to take care of this extra effect; comparing a full
four parameter fit with a three parameter fit having $s_0$ fixed to some
reasonable value shows that even if the continuum threshold is
effectively undetermined the fit for the other parameters is
trustworthy.

Note that in the figures we show only the timeslice zero data, even
though the fit included timeslice one data as well.  This is partly
for clarity (there are fewer points although the shape of the data at
large distances is in all cases indistinguishable to the eye between
the timeslices) and partly to emphasise the failings in our data at
small distances.

The data and fits are shown in figures~\ref{fig:pseudo},
\ref{fig:vector} and~\ref{fig:nucleon}.  It is clear that all the fits
are entirely acceptable bearing in mind our difficulties for $r<6a$.
All the fits have chi-squared per degree of freedom around one or
less, though again the correlations should be borne in mind.  To
enable the reader to make his or her own judgements we have labelled
the axes in lattice units; in physical units, the inverse lattice
spacing is $2.73(5)$ GeV, or $1\mathrm{fm}$ is approximately $14a$.

Before addressing the fits in detail, a few qualitative remarks are in
order. The contrast between the three channels --- pseudoscalar,
vector, and nucleon --- is striking, and in good qualitative agreement
with the expectations of~\cite{Shu93}. For the pseudoscalar channel
(figure~\ref{fig:pseudo}), $R(r)$ rises sharply from unity (note the
vertical scale is somewhat arbitrary, since we cannot fix the
intercept; also note that $R_P(r)$ rises so steeply that we have plotted
the data using a logarithmic scale) at distances of $\sim0.3{\rm fm}$,
showing that asymptotic
freedom is violated sharply in this channel. No doubt this can be
related to the presence of the Goldstone pion mode in this channel
dictated by the spontaneous breakdown of chiral symmetry in the QCD
vacuum. In the vector channel (figure~\ref{fig:vector}), we see that
$R(r)$ is roughly constant, varying by less than 50\%, out to
$r\simeq1{\rm fm}$ (in fact, very nearly the entire useful extent of
the lattice volume) --- which arises from a conspiracy between the
contributions from the continuum, which dominate at short distances,
and the rho meson resonance which takes over at larger
distances. Shuryak refers to the flatness of $R_V(r)$ out to this
range as ``superduality''~\cite{Shu93}.  In the nucleon channel, the
variation is much greater; $R(r)$ rises by a factor of between 3 and 4
at $r\simeq1{\rm fm}$ over its value near the origin.  Recall that in
all cases, over this range of $r$ both $K_{\rm free}$ and $K_{\rm
interacting}$ fall by many orders of magnitude.

\begin{table}[tb]
\renewcommand{\arraystretch}{1.3}
\newlength\descwidth
\begin{tabular}{lrrrp{3cm}}
\hline
Channel & \multicolumn{1}{c}{$M_\mathrm{resonance}$} & 
	\multicolumn{1}{c}{$\mathstrut\sqrt[n]{\lambda}$} & 
	\multicolumn{1}{c}{$\sqrt{s_0}$} &
	Source\\
\hline\hline
Pseudoscalar & $310(30)$ & $450(20)$ & $1,700(400)$ & 4 parameter fit \\
 & $156(10)$ & $440(10)$ & ${} < 1.0$ & Chu et al.\\
 & $138$ & $480$ & 1,300(100) & Phenomenology\\
\hline
Vector & $690(170)$ & $380(60)$ & $1,400(400)$ & 4 parameter fit \\
 & $[798]$ & $404(40)$ & $1,800(340)$ & 3 parameters, fixed mass \\
 & $720(60)$ & $410(20)$ & $1,620(230)$ & Chu et al.\\
 & $780$ & $409(5)$ & $1,590(20)$ & Phenomenology \\
\hline
Axial & $1,300(300)$ & $610(230)$ & ${}<3,000$ & 3 parameter fit \\
 & $1,260$ & $520(20)$ & ${}\gtrsim 1,500$ & Phenomenology \\
\hline
Scalar & none & none & ${}<1,400$ & 1 parameter fit \\
\hline
Nucleon & $1,150(60)$ & $363(13)$ & $1,200(400)$ & 4 parameter fit \\
 & $[1020]$ & $336(38)$ & & 2 parameters, fixed mass \\
 & $950(50)$ & $293(15)$ & ${}<1.4$ & Chu et al.\\
 & $939$ & \multicolumn{1}{c}{?} & 1,440(40) & Phenomenology \\
\hline\hline
\end{tabular}
\caption{Fit parameters extracted from the data with a four parameter
fit (or three parameters where the continuum part is not well
constrained); the result for the normalisation is not shown.  For
comparison results from Chu et al.~\protect\cite{Chu93} and
Shuryak~\protect\cite{Shu93} are also shown.  All quantities are in
MeV; to achieve this, the second data column shows $\protect\sqrt{\lambda}$
for the mesons and $\protect\sqrt[3]{\lambda}$ for the nucleon
(i.e.\ $n=2$ and $n=3$ respectively).}
\label{tab:fit}
\end{table}

We show the corresponding fit parameters (except the normalisation) in
table~\ref{tab:fit} together with a comparison of the previous
results~\cite{Chu93}.  Mostly the agreement is reasonable.  One
noticeable feature is that our rho is too light and our nucleon too
heavy for experiment.  Looking at the fits, one can see that indeed
the vector fit is above the data points and the nucleon fit below.
This suggests fixing the mass in these channels to the value extracted
by conventional techniques~\cite{All93} and then fitting with the
remaining parameters; in the case of the nucleon, we find the fit is
now unstable unless $s_0$ is also fixed.  These revised fits are shown
as a dashed line in the figures.  We retain $\chi^2/\mathrm{d.o.f}$
near to one; as a bonus, we actually achieve a close match to the
vector data in the small distance range where we have not fitted.
Since the mass is the most well-known of the fitted parameters, there
is some merit in preferring the mass fixed parameters in
table~\ref{tab:fit} to the free mass ones as a guide to extracting
other phenomenological parameters.  The errors quoted for the these
fits include a rough estimate of the effects of the uncertainty due to
the statistical errors on the masses.

Thus it is clear from looking at the data both that there is no
contradiction with the previously extracted mass values and that this
method is not well optimised for extracting them.  The peak in $R(r)$
due to the delta-function in equation~(\ref{eqn:fform}) occurs
just at the point where the data becomes noisy.  This suggests that a
slightly larger physical lattice would have been better for our
particular purposes. We should stress that in this work we only use
data from 2 timeslices, compared to the 48 available; it is not
surprising that our mass predictions are inferior to those of
ref.~\cite{All93}.

One point about the nucleon is that the $\lambda$ parameter is
particularly highly constrained by the fit, as may be seen on the
small error on the original fit; with the fixed mass fit, this is even
more so although now the error on the mass has been added in.
Presumably systematic errors which we have not attempted to estimate
become important for this quantity.


The pseudoscalar, however, cannot be forced into line.  Our pion mass
is non-zero which does not fit with the original chiral extrapolation
to zero pion mass.  Even forcing a fit with the physical pion mass
produces a fit with $\chi^2/\mathrm{d.o.f.}\approx 4$ with some 500
(notional) degrees of freedom.  This sheds some doubt on the
usefulness of the extrapolation for our pseudoscalar.  Three quark
masses is not optimal for any extrapolation and in this case it may be
that subtle small mass effects make a significant difference.  In any
case, it is no surprise that the extrapolation of a pseudo-Goldstone
mode to the chiral limit causes the most problems.

We have also studied the two other light meson channels.  In the axial
channel there are two contributing resonances: the $a_1$
at intermediate distances and a pion resonance at
larger distances which appears with an opposite sign due to kinematic
effects.  In this case a logarithmic extrapolation is inappropriate;
we have used a quadratic extrapolation since there is no reason to
suppose the linear extrapolation is any better in this channel than
those where we can make a direct comparison.

We see the decrease in the ratio due to the pion effects in the axial
channel, but again only in the region where the data is noisy.  This
means that we are unable to fit for the relevant parameters, the pion
mass and the decay constant $f_\pi$.  We therefore fix these to
standard values and examine only the $a_1$ and continuum
contributions.  The fit for the continuum contribution is also poor,
with errors comparable to the central value, so we quote a plausible
but not very trustworthy two standard deviation upper limit and use a
three parameter fit for the remaining parameters.

In the scalar channel there are no resonances and we have only a
normalisation and the continuum contribution.  Fitting the latter is
unstable and again we are only able to quote an upper limit; here it
is chosen such that the $\chi^2/\mathrm{d.o.f.}$ for the remaining
parameters becomes unfeasibly large.  There is a clear region in $s_0$
where this begins to happen; our results definitely favour a continuum
cut-off no more than about $1.5\mathrm{GeV}$.  Figure~\ref{fig:scalar}
shows a fit with $\sqrt{s_0}$ fixed at $0.3a^{-1}$, corresponding to
$270\mathrm{MeV}$, which passes near the central values of the data at
large distances, but clearly this value is not remotely trustworthy.

\section{Conclusion}

In many respects, this has been an exploratory calculation, making use
of existing data designed for other purposes \cite{All93}, but we have
learned many things. First of all, to the accuracy we are able to
achieve, once both statistical and systematic errors are accounted for,
the ansatz (5) appears to be perfectly well able to account for our
results in all the channels examined. Apart from particle masses, which
are best extracted by standard methods, the numerical results from fits
to the data are consistent with those published in the pioneer lattice
study \cite{Chu93}, and also with those derived from phenomenology
\cite{Shu93}, although our errors are in most cases larger. The reason
for this has been firstly our difficulties in exploring the $r\to0$
limit due to the strange behaviour of SW fermions, and secondly
uncertainties in the chiral extrapolation due to the use of just three
$\kappa$ values.

Clearly any subsequent study must improve on both of these problems: in
the first case either via a better analytic understanding of the SW
approach, or by using a different fermion formulation; in the second
case by using more $\kappa$ values. Another possible improvement would
be to use data from even more timeslices in an effort to beat down
statistical errors in the large $r$ regime --- all the while taking care
to check that possible errors due to working on asymmetric lattice
volumes (which are now standard in large-scale QCD simulations)
are under control.

In the future, if improvements prove possible, we can contemplate a more
refined analysis of the data, and in particular a search for evidence
that the ansatz (5) is ultimately insufficient. One possibility would be
to subtract the known resonance peak from the data to focus on the
continuum contribution; another is to attempt to calculate $f(s)$
directly by some numerical inversion of equation (4). It may well prove
necessary to go beyond the simplest model of the data in order to
understand the effects of introducing dynamical quark loops.
An important prerequisite for such a study, however, will be a
systematic understanding of effects due to non-zero $a$ and finite $V$.

Finally, on the phenomenological side, we can do little better than to
re-refer the reader to Shuryak's review \cite{Shu93}. His point of view
is that the hadronic correlation functions studied here are effective
probes of the vacuum structure of QCD, yielding information which
complements measurements of condensates, say. Inspection of our results
for the various channels (figures \ref{fig:pseudo},\ref{fig:vector},
\ref{fig:nucleon}) reveals an obvious contrast between the pseudoscalar
channel, where departure from the non-interacting theory begins at very
short distances $\sim0.3{\rm fm}$, and the vector and nucleon channels,
which remain reasonably close to the non-interacting prediction for
distances out to 1fm. Perhaps this may be taken as evidence that the two
crucial non-perturbative features of QCD, chiral symmetry breaking and
colour confinement, are due to mechanisms taking place at different energy
scales.

\section*{Acknowledgments}
This work was supported by an EC grant number ERBCHRXCT920051.
SJH was supported
by a PPARC Advanced Fellowship.  We thank David Henty for providing
the free fermion propagators used for testing, and Jeff Grandy for
information on the method of performing the image corrections.


\begin{thebibliography}{99}
\bibitem{Shu93} E. Shuryak, Rev.\ Mod.\ Phys.\ 65 (1993) 1.
\bibitem{Chu93} M.-C. Chu, J.M. Grandy, S. Huang and J.W. Negele,
Phys.\ Rev.\ D48 (1993) 3340.
\bibitem{Kro93} A.S. Kronfeld and P.B. Mackenzie, `Progress in QCD
using lattice gauge theory', Ann.\ Rev.\ Nucl.\ Part.\ Phys.\ 43
(1993) 793.
\bibitem{All93} UKQCD collaboration, C.R. Allton et al.,
Nucl.\ Phys.\ B407 (1993) 331
\bibitem{She85} B. Sheikholeslami and R. Wohlert, Nucl.\ Phys.\ B259
(195) 572.
\bibitem{Hea91} G. Heatlie, C.T. Sachrajda, G. Martinelli, C. Pittori
and G.C. Rossi, Nucl.\ Phys.\ B352 (1991) 266.
\bibitem{Bur94} M. Burkardt, J.M. Grandy and J.W. Negele, `Calculation
and interpretation of hadron correlation functions in lattice QCD',
MIT preprint MIT-CTP-2109, hep-lat/9406009
\bibitem{Car85} D.B. Carpenter and C.F. Baillie, Nucl.\ Phys.\ B260
(1985) 103.
\end{thebibliography}
\end{document}